\documentclass[preprint,aps,secnumarabic,showpacs,nobibnotes,amsmath,amssymb,prd]{revtex4}
\newcommand{\be}{\begin{equation}}
\newcommand{\ee}{\end{equation}}
\newcommand{\bq}{\begin{eqnarray}}
\newcommand{\eq}{\begin{eqnarray}}
\preprint{}
\begin{document}
\title{\large\bf Supersymmetric Standard Model from String Theory}
\medskip
\author{B. B. Deo }
\affiliation{ Physics Department, Utkal University, Bhubaneswar-751004, India.}
\begin{abstract}
N=1, D=4 Superstring possessing a $SO(6)\otimes SO(5)$ symmetric action  
 and with the same  gauge symmetry obtained from the zero mass spectrum of vector meson
as well, is constructed from the bosonic string in twenty six dimensions. Without 
breaking supersymmetry, the gauge symmetry of the model descends to the supersymmetric 
standard model of the electroweak scale in four  dimension. It is proved that there can be three 
generations in the model.
\end{abstract}
\maketitle

The model begins from the Nambu-Goto~\cite{Nambu70, Goto71} bosonic string theory in the world sheet 
($\sigma ,\tau$) which makes sense in 26 dimensions. The reason is easy to see. The string 
action in twenty six dimension is
\begin{equation}
S_B=-\frac{1}{2\pi}\int d^2\sigma 
\left ( \partial_{\alpha}X^{\mu}\partial^{\alpha} X_{\mu} \right ),~~~\mu =0,1,2,...,25.\label{e1}
\end{equation}
where $\partial_{\alpha}=(\partial_{\sigma}, \partial_{\tau})$. The central charge for bosons is
easily found by using the general expression for the two energy momentum tensors at two world sheet points z
and $\omega$
\begin{equation}
2< T(z)T(\omega) > = \frac{C}{(z-\omega)^4} + .....\label{e2}
\end{equation}
The coefficient of the most divergent term as C in equation (\ref{e2}) is the central charge. For free
bosons, the central charge is
\begin{equation}
C_B=\delta_{\mu}^{\mu}\label{e3}
\end{equation}
For equation (\ref{e1}) i.e. the action $S_B$ has a central charge $\delta_{\mu}^{\mu}$=26. So the
action (\ref{e1}) is not anomaly free. One adds to equation (\ref{e1}) of the action of the
conformal ghosts ($c^+, b_{++}$)
\begin{equation} 
S_{FP}=\frac{1}{\pi} \int\left ( c^+\partial_-b_{++} + c^-\partial_+c_{--}\right ) d^2\sigma \label{e4}
\end{equation}
This action (\ref{e4}) has  a central charge $-$26, independent of the dimensionality of the string. To have an
anomaly free string theory, the central charge of the conformal ghosts should be able to cancel only when
the C=$\delta^{\mu}_{\mu}$=D=26. The string is physical only in D=26 dimension with the 
total central charge zero.
Using Mandelstam's~\cite{Mandelstam75} proof of equivalence between one boson to two fermionic 
modes in the infinite volume limit in (1+1) dimensional quantum field theory, one can rewrite the action 
as the sum of the four bosonic coordinates $X^{\mu}$ of SO(3,1) 
and forty four fermions having internal symmetry SO(44). If this is anomaly free, this is also true 
in finite intervals or a circle. Noting that the Majorana fermions can be in bosonic representation of 
the Lorentz group SO(3,1), 
the forty four fermions are grouped into eleven Lorentz vectors of SO(3,1) which look as a commuting
internal symmetry group when viewed from the other internal quantum number space.
The action is now
\begin{equation}
S_{FB}=-\frac{1}{2\pi}~\int~d^2 \sigma \left [ \partial^{\alpha}X^{\mu}(\sigma,\tau)~
\partial_{\alpha}X_{\mu}(\sigma,\tau) - i\sum_{j=1}^{11}\bar{\psi}^{\mu,j} 
\rho^{\alpha}\partial _{\alpha}\psi_{\mu,j} \right ].\label{e5}
\end{equation}
and is anomaly free with $S_{FP}$ of equation (\ref{e4}).
The upper indices j,k refer to a row and that of lower to a column and
\begin{eqnarray}
\rho^0 =
\left (
\begin{array}{cc}
0 & -i\\
i & 0\\
\end{array}
\right )
\end{eqnarray}
and
\begin{eqnarray}
\rho^1 =
\left (
\begin{array}{cc}
0 & i\\
i & 0\\
\end{array}
\right )
\end{eqnarray}
Dropping indices
\begin{equation}
\bar{\psi}=\psi^{\dag}\rho^0.
\end{equation}
$\rho^{\alpha}$'s are imaginary so the Dirac operators $\rho^{\alpha}\partial_{\alpha}$ is real and 
this representation of Dirac algebra, the components of the world sheet spinor $\psi^{\mu,j}$ are
real and are Majorana spinors.

One has introduced an anticommuting field $\psi^{\mu,j}$ that transforms as vectors - a bosonic 
representation of SO(3,1). $\psi^{\mu}_A$ maps bosons to bosons and fermions to fermions in the 
space time sense. There is no clash with spin statistics theorem. Action (\ref{e5}) is a two dimensional
field theory, not a  field theory is space time. $\psi^{\mu}_A$ transforms as a spinor under the 
transformation of the two dimensional world sheet. The Lorentz group SO(3,1) is merely an internal symmetry
group as viewed in the world sheet. This is discussed in reference \cite{Green87}.

The central charge of the the free fermions in action (\ref{e5}) as deduced by calculation from equation
(\ref{e2}) is
\begin{equation}
C_F= \frac{1}{2} \delta^{\mu}_{\mu}\delta_j^j\label{e1a},
\end{equation}
so that the total central charge of equation (\ref{e5}) is
\begin{equation}
C_{ SB}=\delta_{\mu}^{\mu} + \delta_{\mu}^{\mu}\delta_j^j = 4+ \frac{1}{2}\times 4\times 11=26\label{e6}
\end{equation}
The total central charge is zero with $S_{FP}$. As a check of equation (\ref{e6}), the central charge of a
ten dimensional superstring is 10+$\frac{1}{2}\times 10\times 1=15$ . So this formula is correct. 
However, the action 
(\ref{e5}) is not yet supersymmetric. The eleven $\psi^{\mu,j}_A$ have to be further 
divided into two species; $\psi^{\mu,j}$, j=1,2,..6 and $\phi^{\mu, k}$, $k$=7,8..11. 
For the group of six, the positive and negative parts are $\psi^{\mu,j}=\psi^{(+)\mu,j} +\psi^{(-)\mu,j}$  
whereas for the group of five, allowed the freedom of phase of creation operators for Majorana fermions
in $\phi^{\mu,k} = \phi^{(+)\mu,k}-\phi^{(-)\mu,k}$. The action is now

\begin{equation}
S= -\frac{1}{2\pi} \int d^2\sigma\left [\partial_{\alpha}X^{\mu}\partial^{\alpha}X_{\mu}
-i~\bar{\psi}^{\mu,j}~\rho^{\alpha}~\partial_{\alpha}~\psi_{\mu,j}
+ i~\bar{\phi}^{\mu,k}~\rho^{\alpha}~\partial_{\alpha}~\phi_{\mu,k}\right ]
\label{e7}
\end{equation}

Besides SO(3,1), the action (\ref{e7}) is invariant under $SO(6)\otimes SO(5)$. It is also 
invariant under the supersymmetric transformation
\begin{eqnarray}
\delta X^{\mu} =\bar{\epsilon}~(e^j\psi^{\mu}_j - e^k\phi^{\mu}_k),\\
\delta\psi^{\mu,j}= - ie^j\rho^{\alpha}\partial_{\alpha}X^{\mu}~\epsilon\\
\delta\phi^{\mu,k}= ie^k\rho^{\alpha}\partial_{\alpha}X^{\mu}~\epsilon.\label{e8}
\end{eqnarray}
$\epsilon$ is a constant anticommuting spinor. $e^j$ and $e^k$ are taken as eleven numbers of a 
row matrix with $e^j e_j$=6 and $e^k e_k$=5. They look like~ $e^1$=(1,0,0,0,0,0,0,0,0,0,0) ~and
~$e^7$=(0,0,0,0,0,0,1,0,0,0,0).

There is wide mismatch between the fermionic and bosonic modes in the action (\ref{e7}).
To investgate these disturbing features, we find that the commutators of two successive 
supersymmetric transformations 
lead to a translation with the coefficients $a^{\alpha} = 2~i~\bar{\epsilon}^1
\rho^{\alpha}\epsilon_2$ provided the internal symmetry indices (SO(11) in equation (\ref{e5}) or
SO(6)$\otimes$SO(5) in equation(\ref{e7})) satisfy.
\begin{equation}
\psi_j^{\mu} = e_{j}\Psi^{\mu},~~~~~~~~~\phi_k^{\mu} = e_{ k}\Psi^{\mu},\label{e9}
\end{equation}
These are the two key equations. They also state that of the eleven sites (j,k), the super fermionic partner
$\Psi^{\mu}$ is found in one site only.
On  j$^{th}$ or k$^{th}$ site, it emits or absorbs quanta as found by quantising $\psi^{\mu,j}$ or
$\phi^{\mu,k}$ respectively prescribed by the action (\ref{e7}). The alternative auxilliary 
fields are not needed. The $\Psi^{\mu}$ is given by 
\begin{equation}
\Psi^{\mu} = e^j\psi^{\mu}_j - e^k\phi^{\mu}_k.\label{e10}
\end{equation}
It is easy to verify that there is no $e^j$ or $e^k$ and 
\begin{eqnarray}
\delta X^{\mu}=\bar{\epsilon}\Psi^{\mu}, \;\;\;\;\;\;\;\;\; \delta \Psi^{\mu}=-i\;
\epsilon\;\rho^{\alpha}\;\partial_{\alpha}\;X^{\mu}\label{e11}
\end{eqnarray}
and 
\begin{equation}
[\delta_1 ,\delta_2]X^{\mu } = a^{\alpha}\partial_{\alpha}X^{\mu },\;\;\;\;\;\;
[\delta_1 ,\delta_2]\Psi^{\mu } = a^{\alpha}\partial_{\alpha}\Psi^{\mu }\label{e12}
\end{equation}
We immediately obtain a Nambu-Goto superstring in four dimensions from the action (\ref{e7})
using equation (\ref{e9}). The action is
\begin{equation}
S=-\frac{1}{2\pi}~\int~d^2\sigma\left ( \partial_{\alpha} X^{\mu}\partial^{\alpha}X_{\mu}
-~i\Psi^{\mu}\rho^{\alpha}\partial_{\alpha}\Psi_{\mu}\right )\label{e13}
\end{equation}
Thus the action(\ref{e7}) is truely supersymmetric. Quantising this action in well written 
down procedure, the particle spectrum is very rich unlike quantising (\ref{e13}).
We also need the internal symmetry $SO(6)\otimes SO(5)$ of the action (\ref{e7}).

From equations (\ref{e11}) and (\ref{e12}), it follows that the superpartner of $X^{\mu}$
is $\Psi^{\mu}$. Introducing another supersymmetric pair the Zweibein 
$e^{\alpha}(\sigma,\tau)$ and the gravitons $\chi_{\alpha}= \nabla_{\alpha}\epsilon$,
the local 2-d supersymmetric action, first written down by Brink, Di Vecchia, Howe, Deser
and Zumino~\cite{Brink76, Deser76} is
\begin{equation}
S= -\frac{1}{2\pi}\int d^2\sigma ~e~\left [ h^{\alpha\beta}\partial_{\alpha}X^{\mu }
\partial_{\beta}X_{\mu } -i\bar \Psi^{\mu}\rho^{\alpha}\partial_{\alpha}
\bar \Psi_{\mu}+ 2\bar{\chi}_{\alpha}\rho^{\beta}\rho^{\alpha}\Psi^{\mu}
\partial_{\beta}\chi^{\mu}+\frac{1}{2}
\bar{\Psi }^{\mu}\Psi_{\mu}\bar{\chi}_{\beta} \rho^{\beta}\rho^{\alpha}\chi_{\alpha}
\right ]\label{e14}
\end{equation}
This action has several invariances. For detailed discussions, see reference~\cite{Green87}.
Varying the field and Zweibein, vanishing of the Noether current $J^{\alpha}$ and 
the energy momentum tensor $T_{\alpha\beta}$ is derived.
\begin{equation}
J_{\alpha}= \frac{\pi}{2e}\frac{\delta S}{\delta \chi^{\alpha}}=\rho^{\beta}
\rho_{\alpha}\bar{\Psi}^{\mu}
\partial_{\beta}X_{\mu}=0
\end{equation}
and
\begin{equation}
T_{\alpha\beta}=\partial_{\alpha}X^{\mu }
\partial_{\beta}X_{\mu }- \frac{i}{2}\bar{\Psi}^{\mu}\rho_{(\alpha}\partial_{\beta )}
\Psi_{\mu}=0.
\end{equation}
In a light cone basis, the vanishing of the light cone components are
\begin{equation}
J_{\pm}=\partial_{\pm}X_{\mu}\Psi^{\mu}_{\pm}=0\label{e14a}
\end{equation}
and
\begin{equation}
T_{\pm\pm}=
\partial_{\pm}X^{\mu}\partial_{\pm}X_{\mu}+\frac{i}{ 2}\psi^{\mu j}_{\pm}\partial_{\pm}
 \psi_{\pm\mu,j }- \frac{i}{2}\phi_{\pm}^{\mu k}\partial_{\pm}\phi_{\pm\mu,k},\label{e15}
\end{equation}
where $\partial_{\pm}=\frac{1}{2}(\partial_{\tau} \pm\partial_{\sigma})$. 

The central charge of the action (\ref{e7}) is $\text{C} = \delta^{\mu}_{\mu}+\frac{1}{2}
\delta^{\mu}_{\mu}(\delta^j_j+\delta^k_k)$=26 also. This is a supersymmetric action and with the
conformal ghosts, the anomaly vanishes. But there  will be current generator constraints for which
we need superconformal ghosts for BRST charge. I shall come back to this problem later in this article.

The action in equation (\ref{e14}) is not space time supersymmetric. However, in the 
fermionic representation SO(3,1), fermions are Dirac spinors with four components $\alpha$. 
We construct Dirac spinor like equation (\ref{e10}) as the sum of component spinors
\begin{equation}
\Theta_{\alpha}=\sum^6_{j=1}e^j\theta_{j\alpha} -
\sum^{11}_{k=7}e^k\theta_{k\alpha}.
\end{equation}
With the usual Dirac matrices $\Gamma^{\mu}$, since the identity
\begin{equation}
\Gamma_{\mu}\psi_{[1}\bar{\psi}_2\Gamma^{\mu}\psi_{3]} =0
\end{equation}
is satisfied due to the Fierz transformation in four dimension, the Green-Schwarz action
~\cite{Green84} for N=1 supersymmetry is
\begin{equation}
S=\frac{1}{2\pi}\int d^2\sigma \left ( \sqrt{g}g^{\alpha\beta}\Pi_{\alpha}\Pi_{\beta}
+2i\epsilon^{\alpha\beta}\partial_{\alpha}X^{\mu}\bar{\Theta}\Gamma_{\mu}
\partial_{\beta}\Theta\right ),\label{e16}
\end{equation}
where 
\begin{equation}
\Pi^{\mu}_{\alpha}=\partial_{\alpha}X^{\mu}- i\bar{\Theta}\Gamma^{\mu}\partial_
{\alpha}\Theta .\label{e17}
\end{equation}

This is the N=1 and D=4 superstring originating from the D=26 bosonic string. It is 
difficult to quantise this action covariantly. It is better to use NS-R~\cite{Neveu71, Ramond71} 
formulation with G.S.O projection~\cite{Gliozzi76}. This has been done in reference~\cite{Deo03, Deo103},
while quantising each eleven component fermions $\psi_j^{\mu}$ and $\phi_k^{\mu}$, there appears eleven
additional ghosts, $\eta^{00}$ being -1. But, there can be similarly $11\times2$ constraint equations 
from one equation (\ref{e14a}). Since the superpartner $\Psi^{\mu}$ has been identified, we omit further 
refering to $e_j$, $e_k$ for easy reading.

To outline briefly, let $L_m$, $G_r$ and $F_m$ be the Super Virasoro generators of energy, 
momenta and currents. Let $\alpha$'s denote the quanta of $X^{\mu}$ field, b's and 
b$^{\prime}$'s denote quanta of $\psi$ and $\phi$ fields in NS formulation and d, d$^{\prime}$ in 
R formulation. Then,
\begin{eqnarray}
L_m&=&\frac{1}{\pi}\int_{-\pi}^{\pi}d\sigma e^{im\sigma}T_{++}\nonumber\\
 &= &\frac{1}{2}\sum^{\infty}_{-\infty}:\alpha_{-n}\cdot\alpha_{m+n}: +\frac{1}{2}
\sum_{r\in z+\frac{1}{2}}(r+\frac{1}{2}m): (b_{-r} \cdot b_{m+r} - b_{-r}' \cdot b_{m+r}'):~~~~~~~~~NS
\nonumber\\
&=&\frac{1}{2}\sum^{\infty}_{-\infty}:\alpha_{-n}\cdot\alpha_{m+n}: +\frac{1}{2}
\sum^{\infty}_{n=-\infty}(n+\frac{1}{2}m): (d_{-n} \cdot d_{m+n} - d_{-n}'
\cdot d_{m+n}'):~, ~~~~~~~R\\
G_r &=&\frac{\sqrt{2}}{\pi}\int_{-\pi}^{\pi}d\sigma e^{ir\sigma}J_{+}=
\sum_{n=-\infty}^{\infty}\alpha_{-n}\cdot \left( e^jb_{n+r,j}- e^kb'_{n+r,k}\right ), ~~~~~~~~~~~~~~~~~~~~~~~~~~~~~~ NS\\
\text{and}\nonumber\\
F_m &=&\sum_{-\infty}^{\infty} \alpha_{-n}\cdot \left( e^jd_{n+r,j}- e^kd'_{n+r,k}\right ) .~~~~~~~~~ 
~~~~~~~~~~~~~~~~~~~~~~~~~~~~~~~~~~~~~~~~~~~~~~~~ R\label{e18}
\end{eqnarray}
and satisfy the super Virasoro algebra with central charge C
\begin{eqnarray}
\left [L_m , L_n\right ] & = &(m-n)L_{m+n} +\frac{C}{12}(m^3-m)\delta_{m,-n},\\
\left [L_m , G_r\right ] & = &(\frac{1}{2}m-r)G_{m+r}, ~~~~~~~~~~~~~~~~~~~~~~~~~~~~~~~~~~~~~~~~~~~~~NS\\
\{G_r , G_s\} & =& 2L_{s+r} +\frac{C}{3}(r^2-\frac{1}{4})\delta_{r,-s},\\
\left [L_m , F_n\right ] & = & (\frac{1}{2}m-n)F_{m+n},~~~~~~~~~~~~~~~~~~~~~~~~~~~~~~~~~~~~~~~~~~~~~~R\\
\{F_m, F_n\} & = & 2L_{m+n} +\frac{C}{3}(m^2-1)\delta_{m,-n},\;\;\;\; m\neq 0.\label{e19}
\end{eqnarray}

This is also known that the normal ordering constant of
$L_o$ is equal to one and we define the physical states as satisfying
\begin{eqnarray}
(L_o-1)|\phi>&=&0,~~~L_m|\phi>=0,~~~G_r|\phi>=0~~~ for~~~~ r,m>0,\;\;\;NS\;\;\;Bosonic\label{e20}\\
 L_m|\psi>&=&F_m|\psi>=0\;\;\;\;\;\;\;\;\;\;\;\;\;\;\;
for\;\;\;\;\; m>0,\;\;\;\;\;\;\;\; ~~:R\;\;\;\;Fermionic\label{e21}\\
(L_o-1)|\psi>_{\alpha}&=&(F_o^2-1)|\psi>_{\alpha}=0.\label{e22}
\end{eqnarray}
So we have
\begin{equation}
(F_o +1)|\psi_+>_{\alpha}=0\;\;\;\; and\;\;\; (F_o-1)|\psi_->_{\alpha}=0~~~~~~~~~~:R\label{e23}
\end{equation}
These conditions shall make the string model ghost free.

It can be seen in a simple way. Applying $L_o$ condition
the state  $\alpha_{-1}^{\mu}|0,k>$ is massless and the $L_1$ constraint gives the
Lorentz condition $k^{\mu}|0,k>=0$ implying a transverse photon and
Gupta Bleuler impose that $\alpha_{-1}^0|\phi>=0$. Applying $L_2,\;L_3\; ....$, constraints,
one obtains $\alpha_{m}^0|\phi>=0$. Further, since $[\alpha_{-1}^0 , G_{r+1}]|\phi>=0$,\;\;
we have $b_{r,j}^0|\phi>=0$ and $b_{r,k}^{'0}|\phi>=0$. All the time components are 
eliminated from Fock space.

The central charge, coming from (\ref{e7}) is 26. This is cancelled by the contribution of the action
of the conformal ghosts -26. However, one must have superconformal ghosts which contributes a
central charge 11 so that the current generators constraints (\ref{e21}) and (\ref{e22}) are 
preserved by the BSRT nilpotent charge. We search for a action which is ghost like and and has
central charge 11. It is easily found. The light cone part of the action (\ref{e7}) namely
\begin{equation}
S_{l.c}=\frac{i}{2\pi}\int d^2\sigma\sum_{\mu =0}^3 \left ( \phi^{\mu,j}\rho^{\alpha}\partial_{\alpha}
\phi_{\mu,j} - \phi^{\mu,j}\rho^{\alpha}\partial_{\alpha}\phi_{\mu,j}\right )\label{e24}
\end{equation}
has central charge 11 using equation (\ref{e1a}). The superconformal ghost action should 
be the same in a supplemented ghost space 
where the fermions behave like bosons and vice versa. A separate action for superconformal ghosts will 
overcount the central charge by 11.

I examine the possibility of the action $S_F^{l.c}$ contained in equation (\ref{e7}) to be 
transformable to the action of superconformal ghosts. It is simpler to use 
$\Psi^{\pm}= \frac{1}{\sqrt{2}}(\Psi^0\pm\Psi^3)$
so that the action is 
\begin{equation}
S_F^{l.c}= -\frac{i}{\pi} \int d^2\sigma~\bar{\Psi}^+\rho^{\alpha}
\partial_{\alpha}\Psi^-\label{e25}
\end{equation}
One supplement the ghost space by the superconformal ghost space obeying reverse statistics.
One can do this by letting $\Psi^{\pm}(\sigma)$ to be dependent on Grassman variables 
$\theta$ and $\overline{\theta}$ as well. Then the action takes the form
\begin{equation}
S_F^{l.c}=- \frac{i}{\pi} \int d^2\sigma~\int d\bar{\theta}\int d\theta~
\bar{\Psi}^+(\bar{\theta})~\rho^{\alpha}\partial_{\alpha}
\Psi^-(\theta).\label{e26}
\end{equation}
Expanding the fields as

\begin{equation}
\bar{\Psi}^+(\bar{\theta})= \cdots + \bar{\theta}\bar{\gamma} +\cdots\label{e99}
\end{equation}

and

\begin{equation}
+2~i~\rho^{\alpha}\Psi^-(\theta) = \cdots  + \theta\beta^{\alpha}+\cdots\label{e100}
\end{equation}

The Superconformal ghost action is pulled out  as
\begin{equation}
S_F^{l.c}= -\frac{1}{2\pi} \int
d^2\sigma~\bar{\gamma}\partial_{\alpha}\beta_{\alpha}\label{e101}
\end{equation}
The energy momentum tensor is
\begin{equation}
T_{++}= -\frac{1}{4}\gamma~\partial_+\beta -\frac{3}{4}\beta\partial_+\gamma,\label{e102}
\end{equation}
and the wave equations are ~~$\partial\gamma = \partial\beta=0$.
This has central charge With `p' integral for Ramond and half integral for NS, the field 
quanta $\gamma_p,~\beta_p$ satisfy the only nonvanishing commutator relation
\begin{equation}
[\gamma_p, \beta_r ]= \delta_{p+r}. \label{e103}
\end{equation}
The $L^{l.c}$-generator is transformed into
\begin{equation}
L_m^{l.c.} = \sum_n ~(n + \frac{1}{2}m) ~~:\beta_{m-n}\gamma_n~:\label{e104}
\end{equation}
This is equal to the superconformal ghosts of the normal 10-d superstring.

All that we need to prove that the nilpotency of BSRT charge is that
the conformal dimension of $\gamma$ is `-1/2' and that of $\beta$ is `3/2' respectively. 
They can be deduced from equation(\ref{e104}) using equation (\ref{e103}).
We now proceed to write the nilpotent BRST charge. The part of the charge which comes 
from the usual conformal Lie algebra technique is
\begin{equation}
(Q_1)^{NS,R} = \sum (L_{-m}c_m)^{NS,R} -\frac{1}{2}\sum (m-n) : c_{-m}c_{-n}b_{m+n}:~ -
~a~ c_0 ;~~~~~Q^2_1=0 ~~~for ~a=1.
\end{equation}

Using the Graded Lie algebra, we get the
additional BRST charge, in a straight forward way, in NS and R,
\begin{eqnarray}
Q_{NS}'&=&\sum G_{-r}\gamma_r -\sum\gamma_{-r}\gamma_{-s}b_{r+s},\nonumber\\
Q_{R}'&=&\sum F_{-m}\gamma_m -\sum\gamma_{-m}\gamma_{-n}b_{n+m}.
\end{eqnarray}
As constructed, the BRST charge
\begin{equation}
Q_{BRST}=Q_1 + Q',
\end{equation}
is such that~~$Q_{BRST}^2=0$~~in both the NS and the R sector~\cite{Deo03}. In
proving $\{Q',Q'\} + 2\{Q_1,Q'\}=0$, we have used the fourier
transforms, the wave equations and integration by parts, and  to show that
\begin{equation}
\sum\sum r^2 \gamma_r\gamma_s\delta_{r,-s}=\sum_r\sum_s\gamma_r\gamma_s\delta_{r,-s}=0.
\end{equation}
Thus the  theory is unitary and ghost free.

Interestingly, Polchinski~\cite{Polchinski98} has proved that such actions
allow only a very limited possible algebra. They are shown in a table depending on
the central charge of the ghosts. The present case belongs to either N=0 with 
$C^{ghost}$=-26 or N=1 with $C^{ghost}$=-15. More interestingly, Polchinski adds that 
for N=0 and N=1, there can also be additional spin 1 and spin $\frac{1}{2}$ constraints,
provided that the supercurrent is neutral under the corresponding symmetry. These 
larger symmetries are not essentially different. The total central charge of the ghosts
 allow additional matter but the additional constraints precisely remove the added states, so that
they reduce to the N=0 or N=1 theories. In our case, we started out for the action (\ref{e7})
with $C_{FP}$=-26 i.e. an N=0 supersymmetry. But due to fermionic matter and the current constraints
(\ref{e20}) to (\ref{e23}), the superconformal ghost with $C^{gh}$=11 had had to be added. 
They were taken out of the action (\ref{e7}) as $S_F^{l.c.}$ which was shown to be the same 
as superconformal ghost action equation (\ref{e104}). Thus , finally one ends up with N=1 
supersymmetry with $C^{gh}$=-15.

There are no harmful effective tachyons in the model eventhough 
\begin{eqnarray}
\alpha'M^2&=& -1, -\frac{1}{2}, 0 , \frac{1}{2}, 1, \frac{3}{2},..........NS\\
\text{and}\\
\alpha'M^2&=& -1, 0 , 1, 2, 3,...................R.
\end{eqnarray}
The G.S.O. projection eliminates the half integral values. The tachyonic self energy of 
bosonic sector $<0|(L_o-1)^{-1}|0>$ is cancelled by $-<0|(F_o+1)^{-1}(F_o-1)^{-1}|0>_R$;
the negative sign being due to the fermionic loop. One can proceed a step further and write down
the world sheet supersymmetric charge
\begin{equation}
Q=\frac{i}{\pi}\int_0^{\pi} \rho^0\rho^{\dag\alpha}\partial_{\alpha}X^{\mu}\Psi_{\mu} d\sigma
\end{equation}
and find, as it must,
\begin{equation}
 \sum\{Q_{\alpha}^{\dag}, Q_{\alpha}\} =2H ~~~~~\text{and} ~~~~~ \sum_{\alpha} |Q_{\alpha}
|\phi_o>|^2 = 2 <\phi_o|H|\phi_o> .
\end{equation}
The ground state is of zero energy. There are no overall tachyons in this Superstring.

To prove the modular invariance, one has to use the G.S.O. condition. In covariant
formulation, the number of degrees of freedom of fermions is the number obtained after substraction
of constraints from the total number. In our case the total number is 44 and there are four
constraints. So the physical fermionic modes is 40. The partition function Z can be found by
putting the 8 such fermions($2^3$ of SO(6)~) in each of five boxes. This is a multiplication of 
spin structure of eight fermions given by,
\begin{equation}
A_8(\tau) = (\Theta_3 (\tau)/\eta(\tau))^{4} -
(\Theta_2 (\tau)/\eta(\tau))^{4}-(\Theta_4(\tau)/\eta(\tau))^{4}
\end{equation}
and
\begin{equation}
A_8(1+\tau)=-A(\tau)=A(-\frac{1}{\tau}),
\end{equation}
where the $\Theta(\tau)$'s are the Jacobi theta function, $q=exp(i\pi\tau)$ and ~~$\eta(\tau)$ is
Didekind eta function. Due to the Jacobi relation $A_8(\tau)=0$, the entire partition function 
which is the product of all constituent partition functions~\cite{Deo103} of the model vanishes.

The zero mass particle spectrum of the quantised action(\ref{e7}) is very large. They are 
scalars, vectors and tensorial in nature. In reference~\cite{Deo04} with Maharana, we have 
shown that the massless excitations of the standard model can be found from this model. There 
are also Graviton and Gravitino~\cite{Deo104}. For the gauge symmetry, one has to find massless vector 
bosons which are generators of a group. We follow the work of Li~\cite{Li72}.

Consider O(n). There are $\frac{1}{2}n(n-1)$ generators represented by 
\begin{equation}
L_{ij}= X_i\frac{\partial}{\partial X_j}-X_j\frac{\partial}{\partial X_i}, ~~i,j=1,....,n.
\end{equation}
Using 
\[ \left [ \frac{\partial}{\partial X_i}, X_j\right ] =\delta_{ij} \],
the Lie algebra which is the commutator relation among the generators
\begin{equation}
\left [ L_{ij}, L_{kl} \right ] = \delta_{jk} L_{il}+\delta_{il} L_{jk}-\delta_{ik} L_{jl}-\delta_{jl} L_{ik}.
\end{equation}
Hence one must have $\frac{1}{2}n(n-1)$vector gauge bosons $W^{\mu}_{ij}$ with the transformation law
\begin{equation}
W^{\mu}_{ij}\rightarrow W^{\mu}_{ij}+\epsilon_{ik} W^{\mu}_{kj}+ \epsilon_{jl}  W^{\mu}_{li}, ~~~~
W^{\mu}_{ij}= -W^{\mu}_{ji},
\end{equation}
where $\epsilon_{ij}= - \epsilon_{ji}$ are the infinitesimal parameters which characterised such rotation in
O(n). Under gauge transformation of second kind,
\begin{equation}
W^{\mu}_{ij}\rightarrow W^{\mu}_{ij}+\epsilon_{ik} W^{\mu}_{kj}+ \epsilon_{jl}  W^{\mu}_{li}+\frac{1}{g}
\partial^{\mu}\epsilon_{ij}.
\end{equation}
The Yang-Mills Lagrangian is then written as
\begin{equation}
L=-\frac{1}{4}|F^{\mu\nu}_{ij}|^2
\end{equation}
with 
\begin{equation}
 F^{\mu\nu}_{ij} = \partial^{\mu}W^{\nu}_{ij}-\partial^{\nu}W^{\mu}_{ij}+g\left (W^{\mu}_{ik}W^{\nu}_{kj}
 -W^{\nu}_{ik}W^{\mu}_{kj}\right ),
\end{equation}
$F^{\mu\nu}_{ij}$ has the obvious properties, namely 
\begin{equation}
\Box F^{\mu\nu}_{ij}=0,~~~~ \partial_{\mu}F^{\mu\nu}_{ij}=\partial_{\nu}F^{\mu\nu}_{ij}=0,
~~~~~~~ F^{\mu\mu}_{ij}=0.
\end{equation}

There are two sets of field strength tensors which are found in the model, 
one for SO(6) and the other
for SO(5). Since $\Box F^{\mu\nu}_{ij}=0 $, one can take plane wave solution and write 
\begin{equation}
F^{\mu\nu}_{ij}(x)=F^{\mu\nu}_{ij}(p)~exp(ipx)
\end{equation}
then
\begin{equation}
p^2 F^{\mu\nu}_{ij}(p)=p_{\mu}F^{\mu\nu}_{ij}(p)=p_{\nu}F^{\mu\nu}_{ij}(p)=
F^{\nu\nu}_{ij}(p)=0,~~~~~~~~~~
F^{\mu\nu}_{ij}(p)=-F^{\mu\nu}_{ji}(p).
\end{equation}
These are physical state conditions (\ref{e20})-(\ref{e22}) as well.
\begin{equation}
L_0 F^{\mu\nu}_{ij}(p)=0,~~~~~~~~G_{\frac{1}{2}} F^{\mu\nu}_{ij}(p)=0,
\text{and}~~~~~~L_1 F^{\mu\nu}_{ij}(p)=0.
\end{equation}

The field strength tensor is found to be
\begin{equation}
F^{\mu\nu}_{ij}(p)= b^{\mu\dag}_i~b^{\nu\dag}_j |0,p>;
\end{equation}
(i,j)=1,...,6~for O(6) and 1,....,5 for O(5) with b$'$ replaced by b.
For simplicity, we drop the $'\text{p}'$  dependance and the $\dag$'s so that
$F^{\mu\nu}_{ij}=b^{\mu}_i~b^{\nu}_j $. 
In terms of the excitation of the quanta of the string, the vector generators are
\begin{equation}
W^{\mu}_{ij}=\frac{1}{\sqrt{2ng}}~n_{\kappa}
\epsilon^{\kappa\mu\nu\sigma}b_{\nu,i} b_{\sigma,j}
\end{equation}
$n_{\kappa}$ is the time like four vector and can be taken as (1,0,0,0). One finds that
\begin{eqnarray}
\partial^{\mu}W^{\nu}_{ij}-\partial^{\nu}W^{\mu}_{ij}
&=&p^{\mu}W^{\nu}_{ij}-p^{\nu}W^{\mu}_{ij}\nonumber\\
&=&\frac{1}{\sqrt{2ng}}\left ( n_{\kappa}\epsilon^{\kappa\nu\lambda\sigma}p^{\mu}
-n_{\kappa}\epsilon^{\kappa\mu\lambda\sigma}p^{\nu}\right )b_{\lambda,i} b_{\sigma,j}=0
\end{eqnarray}
as $\mu$ must equal $\nu$ if $\kappa,\lambda,\sigma$ are the same. Again, 
\begin{eqnarray}
g\left (W^{\mu}_{ik}W^{\nu}_{kj} - W^{nu}_{ik}W^{\mu}_{kj}\right )
&=&\frac{1}{2n}\left (  n_{\kappa}~
\epsilon^{\kappa\mu\lambda\sigma}~b_{\lambda,i} b_{\sigma,k}
~n_{\kappa'}~\epsilon^{\kappa'\nu\lambda'\sigma'}~b_{\lambda',k} b_{\sigma,j}
-\mu\leftrightarrow \nu\right )\nonumber\\
&=&b_i^{\mu}b_j^{\nu}=F_{ij}^{\mu\nu}.
\end{eqnarray}
We have used 
\[
\left \{ b_{\lambda'k}, b_{\sigma j}\right \} =
\eta_{\lambda'\sigma}\delta_{kk}=~n~\eta_{\lambda'\sigma}.
\]
 Since the product of pairs of b and b$'$ commute, the gauge group of 
the action~(\ref{e7}) is the product group $SO(6)\otimes SO(5)$. This is same as the 
symmetry of the action (\ref{e7}).

To descend to the standard model group $SU_C(3)\otimes SU_L(2)\otimes U_Y(1)$, 
it can be normally done by 
introducing Higgs which breaks gauge symmetry and supersymmetry. 
However, one uses the method
of symmetry breaking by using Wilson's lines, supersymmetry remains 
in tact but the gauge symmetry is broken. This Wilson loop is
\begin{equation}
U_{\gamma}= P\;\text{exp}(\oint_{\gamma} A_{\mu}\;dx^{\mu}).
\end{equation}
P represents the ordering of each term with respect to the closed path 
$\gamma$. The details of the closed loops are given towards the end of the article
for each particular case. SO(6)=SU(4)
descends to $SU_C(3)\otimes U_{B-L}(1)$. This breaking can be 
accomplished by choosing one element of $U_0$ of SU(4) such that
\begin{equation}
U_0^2=1.
\end{equation}
This element generates the permutation group $Z_2$. Thus
\begin{equation}
\frac{SO(6)}{Z_2}= SU_C(3)\otimes U_{B-L}(1)
\end{equation}
without breaking supersymmetry. Similarly 
$SO(5)\rightarrow SO(3)\otimes SO(2)=SU(2)\otimes U(1)$. 
We have
\begin{equation}
\frac{SO(5)}{Z_2}=SU(2)\otimes U(1)
\end{equation}
Thus
\begin{equation}
\frac{SO(6) \otimes SO(5)}{Z_2 \otimes Z_2}=
SU_C(3)\otimes U_{B-L}(1)\otimes U_R(1) \otimes SU_L(2)
\end{equation}
making an identification with the usual low energy phenomenology. 
But this is not the  standard model. We have an additional U(1). 
There is an instance in $E_6$ where there 
is a reduction of rank by one. Following the same idea~\cite{Green87}, 
we may take

\begin{equation}
U_{\gamma}=(\alpha_{\gamma}) \otimes \left(\begin{array}{ccc}
\beta_{\gamma}&&\\&\beta_{\gamma}&\\&&
\beta^{-2}_{\gamma}\end{array}\right) \otimes 
\left( \begin{array}{ccc}\delta_{\gamma} &&\\&
\delta^{-1}_{\gamma}&\\&&
\end{array}\right) 
\end{equation} 
$\alpha_{\gamma}^3$ =1 such that $\alpha_{\gamma}$ is the cube 
root of unity. This structre lowers the rank by one. We have
\begin{equation}
\frac{SO(6)\otimes SO(5)}{Z_3}=SU_C(3)\otimes SU_L(2)\otimes  U_Y(1)
\end{equation}
arriving at the supersymmetric standard model.

We can elaborately discuss  the $Z_3$ described by ~\cite{Mishra92},
\begin{equation}
g(\theta_1, \theta_2,\theta_3) =
(\frac{2\pi}{3}-2\theta_1,\frac{2\pi}{3}+\theta_2, 
\frac{2\pi}{3}+\theta_3).
\end{equation}
As promised earlier, for one of the first Wilson loop, the angle integral for 
$\theta_1 =\frac{2\pi}{9}$ to $\frac{2\pi}
{3} - \frac{4\pi}{9} = \frac{2\pi}{9}$, so that the loop 
integral vanishes. $\theta_2 =0$ to
$2\pi -\frac{2\pi}{3} =\frac{4\pi}{3}$ for the second loop described 
by a length parameter R. 
$\theta_3 =0$ to $2\pi -\frac{2\pi}{3}= \frac{4\pi}{3}$ 
for the remaining 
loop with the same R. We take the polar components of the gauge 
fields as non zero constants 
as given below.\\
\[ g A_{\theta_2}^{15} = \vartheta_{15}\] for SO(6) = SU(4) 
for which the diagonal generator $t_{15}$ breaks the symmetry and
\[ g'A_{\theta_3}^{10} =\vartheta'_{10} \] for SO(5), 
the diagonal generator being $t'_{10}$. The generators of both SO(6) 
and SO(5)  are $4\times 4$ matrices. We can write the $Z_3$
group as
\begin{eqnarray}
T = T_{\theta_1}T_{\theta_2}T_{\theta_3}.
\end{eqnarray}
We have $T_{\theta_1}$ =1.
This leaves the unbroken symmetry $SU(3)\times SU(2)$ untouched
\begin{equation}
T_{\theta_2}= exp\left( i~t_{15}\int_o^{\frac{4\pi}{3}}\vartheta_{15} R 
d\theta_2\right )
\end{equation}
$T_{\theta_2} \neq 1$ breaks the SU(4) symmetry.
\begin{equation}
T_{\theta_3}= exp\left( i~t'_{10}\int_o^{\frac{4\pi}{3}}\vartheta'_{10} R 
d\theta_3 \right )
\end{equation}
$T_{\theta_3} \neq 1$ breaks the SO(5) symmetry.
But the remaining product of $Z_3$ is
\begin{equation}
T_{\theta_2} T_{\theta_3} =exp \left(
  i\int_o^{\frac{4\pi}{3}}(\vartheta'_{10}t'_{10} +
\vartheta_{15}t_{15}) R d\theta \right )
\end{equation}

Since~~ $\vartheta_{15}$ and $\vartheta'_{10}$~~ are 
arbitrary constants,
we can choose in such a way that ~~
$t_{15}\vartheta_{15} + t'_{10}\vartheta'_{10}=0, 
\frac{3}{2R}, .......$. The term
in the exponential is zero or multiples of $2\pi i$. 
Thus T= U(1) and equation (29)
is obtained, reducing the rank by one~\cite{Polchinski98}.
 
Thus we have made the successful attempt in 
constructing a N=1, D=4 superstring with the gauge symmetry 
$SO(6)\otimes SO(5)$ which, with the help of Wilson 
lines, descend to the SUSY standard model. 
Since the $Z_3$ has three generators and Euler number 3, 
we get three generations of the standard 
model because $SO(6)\otimes SO(5) \rightarrow Z_3\otimes 
SU_C(3)\otimes SU_L(2)\otimes U_Y(1)$. This
is a very important result. The results of this article deserve immediate
and serious attention of the physicists interested in String theory.

I have profitted from a discussion with Prof L.Maharana 
which is thankfully acknowledged.

\end{document}